\documentstyle[12pt]{article}
\newlength{\extraspace}
\newlength{\extraspaces}
\newcommand{\be}{\begin{equation}
\addtolength{\abovedisplayskip}{\extraspaces}
\addtolength{\belowdisplayskip}{\extraspaces}
\addtolength{\abovedisplayshortskip}{\extraspace}
\addtolength{\belowdisplayshortskip}{\extraspace}}
\newcommand{\ee}{\end{equation}}
\newcommand{\ba}{\begin{eqnarray}
\addtolength{\abovedisplayskip}{\extraspaces}
\addtolength{\belowdisplayskip}{\extraspaces}
\addtolength{\abovedisplayshortskip}{\extraspace}
\addtolength{\belowdisplayshortskip}{\extraspace}}
\newcommand{\ea}{\end{eqnarray}}
\newcommand{\nonu}{\nonumber \\[.5mm]}
\newcommand{\A}{&\!\!\!}
\begin{document}
\thispagestyle{empty}
\begin{flushright}
SIT-LP-04/11 \\
{\tt hep-th/0411242} \\
November, 2004
\end{flushright}
\vspace{7mm}
\begin{center}
{\large{\bf Some systematics in the relation between linear 
and nonlinear global supersymmetry in curved spacetime 
}} \\[20mm]
{\sc Kazunari Shima}
\footnote{
\tt e-mail: shima@sit.ac.jp} \ 
and \ 
{\sc Motomu Tsuda}
\footnote{
\tt e-mail: tsuda@sit.ac.jp} 
\\[5mm]
{\it Laboratory of Physics, 
Saitama Institute of Technology \\
Okabe-machi, Saitama 369-0293, Japan} \\[20mm]
\begin{abstract}
We focus on a nonlinear supersymmetry (NL SUSY) 
in curved spacetime introduced in the superon-graviton model (SGM) 
towards a SUSY composite unified model 
based on SO(10) super-Poincar\'e symmetry, 
and we consider for $N = 1$ SUSY a systematic procedure 
to linearize the NL SUSY. 
By introducing modified superspace translations 
of superspace coordinates 
and their specific coordinate transformations 
both depending on a Nambu-Goldstone fermion, 
we show a homogeneous transformation's law of superfields 
which is important in the relation between linear (L) and NL SUSY. 
Furthermore, as a preliminary to find a L supermultiplet 
which is equivalent to the $N = 1$ NL SUSY SGM multiplet, 
we discuss on the realization of the modified superspace translations 
in the construction of a supergravity-like multiplet 
in the superspace formalism. 
In particular, we find constraints on a torsion 
and a Lorentz transformation parameter in the superspace formalism 
to realize the modified supertranslations. 
\end{abstract}
\end{center}

\newpage

Towards a supersymmetric (SUSY) composite unified model 
based on SO(10) super-Poincar\'e (SP) symmetry, 
a new Einstein-Hilbert (EH) type action 
which describes nonlinear (NL) SUSY invariant 
gravitational interaction of Nambu- Goldstone (NG) fermions 
indicating a spontaneous SUSY breaking (SSB) \cite{FI,O} 
has been proposed as superon-graviton model (SGM) of nature \cite{KS}. 
The EH type action of SGM is defined on an extended curved (SGM) spacetime 
where the coset space Grassmann coordinates of superGL(4R)/GL(4R) 
turning to the NG fermions (superon) 
as the fundamental degrees of freedom (d.o.f.) 
are attached at every Riemann spacetime point. 
Therefore, contrary to the simple EH expression of the SGM action, 
the expansion of the action in terms of graviton 
and NG fermions possesses a very complicated (highly nonlinear) 
and rich structure \cite{ST}, 
which consists of the EH action of general relativity (GR), 
the Volkov-Akulov (VA) action \cite{VA} for a NL realization of SUSY 
and their interactions. 
This fact indicates that the vacuum expectation value (v.e.v.) 
of the SGM action in terms of graviton and the NG fermions 
is different from that of the ordinary EH action of GR. 
In other words, a (geometrical) SSB of the SGM action 
to the interacting system of graviton and the NG fermions 
at the early universe is encoded in our SGM scenario. 
We have shown that the SGM action has at least 
[global NL SUSY] $\otimes$ [local GL(4R)] 
$\otimes$ [local Lorentz] $\otimes$ [global SO(10)] 
invariances as a whole which is isomorphic to SO(10) 
SP symmetry \cite{ST}. 

In order to investigate the low energy physics described by the SGM action, 
it is inevitable to derive a linear (L) supermultiplet 
which is equivalent to the NL SUSY SGM multiplet. 
In flat spacetime, the relation between two realizations of global SUSY, 
the L SUSY \cite{WZ1} and the NL SUSY \cite{VA}, 
was investigated through the linearization of NL SUSY of VA 
by many authors \cite{IK}-\cite{STT2}; 
it was shown that the various L supermultiplets \cite{WZ2,Fa} 
with the SSB \cite{FI} are algebraically equivalent 
to the VA action \cite{VA} of NL SUSY. 

In those works of the linearization of NL SUSY, 
it is important to find SUSY invariant relations 
which express basic fields of the L supermultiplets as composites of NG fermions 
and reproduce L SUSY transformations among them 
under the NL SUSY transformation of NG fermions. 
The globally SUSY invariant relations in flat spacetime can be obtained 
by heuristic or by systematic arguments. 
In particular, the systematic procedure in the superspace formalism, 
which was proposed in \cite{IK,UZ}, is useful 
for passing from actions of the L supermultiplets to the VA action. 
In this procedure, by means of specific supertranslations 
of superspace coordinates depending on the NG fermion field, 
it is crucial to find superfields 
which transform homogeneously according to superspace translations 
under the NL SUSY transformations of VA. 
Constraints which lead to the SUSY invariant relations are obtained from 
the homogeneous transformaiton's law of superfields \cite{IK,UZ,STT1}. 

However, the situation in SGM is rather different from the flat spacetime case, 
for the supermultiplet structure of the linearized theory of SGM remains unknown 
except, e.g., for $N = 1$ SUSY it is expected to be a broken SUSY 
supergravity (SUGRA)-like theory containing graviton 
and a (massive) spin-3/2 field. 
Nevertheless, towards the linearization of NL SUSY in SGM 
and also from the viewpoints of the recent interest in NL SUSY 
in the brane world scenario, 
it is useful to consider the application of 
the systematic procedure for linearizing NL SUSY in \cite{IK,UZ} 
to curved (SGM) spacetime. 

In this letter we study the linearization of $N = 1$ SGM. 
We introduce modified superspace translations of superspace coordinates 
and their specific coordinate transformations 
both depending on the NG fermion, 
and we show that they lead to the homogeneous transformation's law 
of superfields which is important in the relation between L and NL SUSY. 
Furthermore, as a preliminary to find a $N = 1$ L supermultiplet 
which is equivalent to the NL SUSY SGM multiplet, 
we also discuss on the realization of the modified superspace translations 
in the construction of a SUGRA-like multiplet by following the superspace formalism 
of Wess-Zumino (WZ) \cite{WZ3,WB} for $N = 1$ SUGRA \cite{FNFDZ}. 
In particular, we find in this letter constraints on a torsion 
and a Lorentz transformation parameter in the superspace formalism 
to realize the modified supertranslations. 

We start with the brief review of the $N = 1$ NL SUSY SGM multiplet \cite{KS,ST}. 
In SGM, the global NL SUSY transformations of a (Majorana) NG fermion $\psi$ 
and a vierbein $e^a{}_\mu$ generated by a global spinor parametar $\zeta$ 
are given by 
\footnote{
Minkowski spacetime indices are denoted 
by $a, b, ... = 0, 1, 2, 3$ 
and four dimensional curved spacetime indices 
by $\mu, \nu, ... = 0, 1, 2, 3$. 
We use the  Minkowski spacetime metric 
${1 \over 2}\{ \gamma^a, \gamma^b \} = \eta^{ab}= (+, -, -, -)$ 
and $\sigma^{ab} = {i \over 4}[\gamma^a, \gamma^b]$. 
}
\ba
\A \A 
\delta_Q \psi = {1 \over \kappa} \zeta 
- i \kappa (\bar\zeta \gamma^\mu \psi) \partial_\mu \psi, \nonu
\A \A 
\delta_Q e{^a}_\mu 
= 2 i \kappa (\bar\zeta \gamma^\nu \psi) \partial_{[\mu} e^a{}_{\nu]} 
\label{NLSUSY-SGM}
\ea
with $\kappa$ being a constant whose dimension is $({\rm mass})^{-2}$ 
and $\partial_{[\mu} e{^a}_{\nu]} 
= (1/2)(\partial_\mu e^a{}_\nu - \partial_\nu e^a{}_\mu)$. 
The NL SUSY transformations (\ref{NLSUSY-SGM}) satisfy closed off-shell 
commutator algebra, 
\be
[\delta_Q(\zeta_1), \delta_Q(\zeta_2)] = \delta_{\rm GL(4R)}(\Xi^\mu). 
\label{SGMalg}
\ee
In Eq.(\ref{SGMalg}), 
$\delta_{\rm GL(4R)}(\Xi^{\mu})$ means a general coordinate transformation 
with a parameter, 
\be
\Xi^\mu = 2 (i \bar\zeta_1 \gamma^\mu \zeta_2 
- \xi_1^\nu \xi_2^\rho e_a{}^\mu \partial_{[\nu} e^a{}_{\rho]}), 
\ee
where $\xi^\mu = - i \kappa \bar\zeta \gamma^\mu \psi$. 
%
%

Eq.(\ref{NLSUSY-SGM}) induces the general coordinate transformation 
of a unified vierbein $w^a{}_\mu$ which is defined 
in terms of $\psi$ and $e^a{}_\mu$ as 
\be
w^a{}_\mu = e^a{}_\mu + t^a{}_\mu 
\label{u-vierbein}
\ee
with $t^a{}_\mu = - i \kappa^2 \bar\psi \gamma^a \partial_\mu \psi$; 
namely, the $w^a{}_\mu$ transforms as 
\be
\delta_Q w^a{}_\mu 
= \xi^\nu \partial_\nu w^a{}_\mu + \partial_\mu \xi^\nu w^a{}_\nu 
\ee
under Eq.(\ref{NLSUSY-SGM}), 
and it has the role of the vierbein defined 
on an extended curved (SGM) spacetime 
where the NG fermion (superon) d.o.f., $\psi$, 
are attached as coset space {\it Grassmann coordinates} 
(i.e., the fundamental d.o.f.) of superGL(4R)/GL(4R) 
at every Riemann spacetime point inspired by NL SUSY. 
Therefore, as an analogy with the geometrical arguments 
of GR on Riemann spacetime, 
the EH type SGM action in terms of $w^a{}_\mu$ (and its inverse 
\footnote{
The inverse $w_a{}^\mu$ in terms of $e^a{}_\mu$ and $\psi$ 
terminates at $O(t^4)$ as 
$w_a{}^\mu = e_a{}^\mu - e_a{}^\nu t^b{}_\nu e_b{}^\mu 
+ e_a{}^\nu t^b{}_\nu e_b{}^\rho t^c{}_\rho e_c{}^\mu 
- e_a{}^\nu t^b{}_\nu e_b{}^\rho t^c{}_\rho e_c{}^\sigma t^d{}_\sigma e_d{}^\mu 
+ e_a{}^\nu t^b{}_\nu e_b{}^\rho t^c{}_\rho e_c{}^\sigma t^d{}_\sigma e_d{}^\kappa 
t^e{}_\kappa e_e{}^\mu$ 
by means of $\psi^n = 0$ for $n \ge 5$.} 
$w_a{}^\mu$), 
which is invariant under the global NL SUSY transformations (\ref{NLSUSY-SGM}), 
is defined \cite{KS} on the SGM spacetime by 
\be
S_{\rm SGM} = -{c^3 \over {16 \pi G}} \int d^4x \vert w \vert (\Omega + \Lambda), 
\label{SGMaction}
\ee
where $\vert w \vert = \det w^a{}_\mu$, $\Omega$ is a scalar curvature 
defined in terms of $w^a{}_\mu$, and $\Lambda$ is a cosmological constant. 
Although the SGM action (\ref{SGMaction}) appalently has the simple form, 
the expansion of Eq.(\ref{SGMaction}) in terms of $e^a{}_\mu$ and $\psi$ 
possesses a very complicated and rich structure \cite{ST}, 
i.e., it is schematically expressed as 
\ba
S_{\rm SGM} \A = \A S_{\rm EH} 
+ (\ S_{\rm VA} + [\ {\rm higher\ derivative\ terms\ of\ the\ NG\ fermion}\ ]\ )
\nonu
\A \A + [\ {\rm interactions\ of\ graviton\ and\ the\ NG\ fermion}\ ], 
\label{SGMexpansion}
\ea
where $S_{\rm EH}$ means the ordinary EH action of GR 
and $S_{\rm VA}$ is the VA action \cite{VA} for the NL realization of SUSY 
in flat spacetime given by 
\ba
S_{\rm VA}(\psi) = \A \A - {1 \over \kappa^2} 
\int d^4 x \ \vert w_{\rm VA} \vert \nonu
= \A \A 
- {1 \over \kappa^2} \int d^4 x 
\left[ 1 + t{^a}_a 
+ {1 \over 2}(t{^a}_a t{^b}_b - t{^a}_b t{^b}_a) \right. \nonu
\A \A 
\left. - {1 \over 3!} \epsilon_{abcd} \epsilon^{efgd} t{^a}_e t{^b}_f t{^c}_g 
- {1 \over 4!} \epsilon_{abcd} \epsilon^{efgh} t{^a}_e t{^b}_f t{^c}_g t{^d}_h 
\right] 
\ea
with $\vert w_{\rm VA} \vert = \det (\delta^a_b + t^a{}_b)$ 
and $t^a{}_b = - i \kappa^2 \bar\psi \gamma^a \partial_b \psi$. 
Eq.(\ref{SGMexpansion}) indicates 
that the v.e.v. of the SGM action 
in terms of graviton and the NG fermions 
is different from that of the ordinary EH action of GR. 
In other words, a (geometrical) SSB of the SGM action (\ref{SGMaction}) 
to the interacting system of graviton and the NG fermions 
at the early universe is encoded in our SGM scenario. 
Also, it can be understood from $e^a{}_\mu \rightarrow 0$ 
in Eq.(\ref{SGMaction}) that the $\Lambda$ 
is a {\it small} cosmological constant related to the strength 
of the superon-vacuum coupling constant $\kappa$ as 
\be
\kappa^2 = ({c^3 \Lambda \over 16 \pi G})^{-1} 
\ee
which is just the fundamental volume in four dimensional spacetime 
of the VA model and gives the mass scale $1/\sqrt{\kappa}$ for the SSB 
\cite{STT1,STT2}. 

Besides the NL SUSY defined by Eq.(\ref{NLSUSY-SGM}), 
the SGM action (\ref{SGMaction}) is invariant 
\footnote{
For $N = 10$ SUSY, the action (\ref{SGMaction}) 
also has the global SO(10) symmetry.}
under the ordinary local GL(4R) 
and the following local Lorentz transformations 
on $w^a{}_\mu$, 
\be
\delta_L w^a{}_\mu = \epsilon^a{}_b w^b{}_\mu 
\ee
with the local parameter $\epsilon_{ab} = \epsilon_{[ab]}(x)$    
or equivalently on $\psi$ and $e^a{}_\mu$
\ba
\A \A \delta_L \psi = - {i \over 2} \epsilon_{ab} \sigma^{ab} \psi, 
\nonu
\A \A \delta_L e^a{}_\mu = \epsilon^a{}_b e^b{}_\mu 
- {\kappa^2 \over 4} \epsilon^{abcd} 
\bar\psi \gamma_5 \gamma_d \psi \ \partial_\mu \epsilon_{bc}, 
\ea
which form a closed algebra \cite{ST}. 

Here we note that the NG fermion (superon) coordinate d.o.f. 
can be gauged away neither by the ordinary GL(4R) 
transformations nor by the local spinor translation, 
e.g., $\delta_\eta \psi(x) = \eta(x)$, 
$\delta_\eta e^a{}_\mu(x) = i \kappa^2 
(\bar\eta(x) \gamma^a \partial_\mu \psi(x) 
+ \bar\psi \gamma^a \partial_\mu \eta(x))$, 
for $w^a{}_\mu(e, \psi)$ of Eq.(\ref{u-vierbein}) 
is an invariant quantity under this transformation 
(for example, see Ref.\cite{STS} for further details). 

Towards the linearization of NL SUSY (\ref{NLSUSY-SGM}) 
to find a L supermultiplet with the spontaneously broken ${\it global}$ SUSY 
which is equivalent to the NL SUSY SGM action (\ref{SGMaction}), 
we consider the application of the sytematic procedure 
for linearizing NL SUSY in \cite{IK,UZ} 
to the curved (SGM) spacetime. 
Although the linearized theory of SGM is unknown, 
it is expected to be a broken SUSY SUGRA-like 
theory containing graviton and a (massive) spin-3/2 field for $N = 1$ SUSY. 
Therefore, we assume in this letter that \\
(a) graviton is not composite (at the leading order, at least) 
of the NG fermion corresponding to 
the vacuum of a Clifford algebra in both NL and L theories \\
and \\
(b) the NL SUSY SGM multiplet ($e^a{}_\mu$, $\psi$) 
should be connected to a L SUSY composite supermultiplet 
($e^a{}_\mu$, $\lambda_\mu(e, \psi)$, 
auxiliary fields($e, \psi$)) 
with $\lambda_\mu$ being a massive spin-3/2 (Majorana) field. 

Let us define a superspace parameterized by superspace 
coordinates $(x^\mu, \theta)$, 
where $\theta$ means a Grassmann coordinate. 
We also denote superfields by $\Phi(x^\mu, \theta)$, 
which is supposed to be, e.g., those used in the superspace formalism 
of $N = 1$ SUGRA (for example, see \cite{WZ3,WB,OS}) 
describing the L supermultiplet 
of ($e^a{}_\mu$, $\lambda_\mu$, auxiliary fields). 
In the linearization of NL SUSY in flat spacetime \cite{IK,UZ}, 
it is crucial to find superfields which transform homogeneously 
according to the ordinary superspace translations 
under the NL SUSY transformations of VA \cite{VA}. 
In the same way, in order to find superfields 
which transform homogeneously according to a certain superspace translation 
under the NL SUSY transformations (\ref{NLSUSY-SGM}), 
we first introduce the following modified superspace translations 
of $(x^\mu, \theta)$ depending on the NG fermion, 
which are generated by a spinor parameter $\epsilon$, 
\ba
\A \A x'^\mu = x^\mu + i \bar\epsilon \gamma^\mu \theta 
- \kappa^2 \bar\theta \gamma^\nu \psi 
\ \bar\epsilon \gamma^\rho \psi \ e_a{}^\mu \nabla_\nu e^a{}_\rho, 
\nonu
\A \A 
\theta' = \theta + \epsilon, 
\label{newstr}
\ea
where $\nabla_\mu$ is the covariant derivative with respect to the spacetime indices, 
i.e., $\nabla_\mu e_a{}^\nu = \partial_\mu e_a{}^\nu 
+ \{^\nu_{\rho \mu} \} e_a{}^\rho$ with the Christoffel symbol $\{^\nu_{\rho \mu} \}$. 
Note that Eq.(\ref{newstr}) reduces to the ordinary superspace translations 
in flat spacetime by taking the limit of $e^a{}_\mu \rightarrow \delta^a_\mu$. 
Here we also mention that the NL SUSY transformation of $\psi$ in Eq.(\ref{NLSUSY-SGM}) 
is derived from Eq.(\ref{newstr}) on a hypersurface in the superspace defined by 
$\theta = \kappa \psi(x)$; namely, the superspace translations (\ref{newstr}) 
on $\theta = \kappa \psi(x)$ by the global spinor parameter $\zeta$ become 
\ba
\A \A x'^\mu = x^\mu + i \kappa \ \bar\zeta \gamma^\mu \psi(x), 
\nonu
\A \A 
\psi' = \psi + {1 \over \kappa} \zeta, 
\ea
which mean $\delta_Q \psi$ in Eq.(\ref{NLSUSY-SGM}). 

Replacing the parameter $\epsilon$ in Eq.(\ref{newstr}) 
by $- \kappa \psi(x)$, we second define specific supertranslations as 
\ba
\A \A x'^\mu = x^\mu - i \kappa \ \bar\psi \gamma^\mu \theta, 
\nonu
\A \A 
\theta' = \theta - \kappa \psi, 
\label{newcd}
\ea
which are just a curved spacetime version of the specific supertranslations 
introduced in \cite{IK,UZ}. 
Then we can prove that superfields defined on $(x'^\mu, \theta')$ of Eq.(\ref{newcd}) 
transform homogeneously according to the modified superspace translations (\ref{newstr}) 
by the global spinor parameter $\zeta$ 
under the NL SUSY transformations (\ref{NLSUSY-SGM}); 
indeed, when we denote superfields on $(x'^\mu, \theta')$ by 
\be
\Phi(x'^\mu, \theta') 
= \tilde \Phi(x^\mu, \theta; e^a{}_\mu(x), \psi(x)), 
\ee
the superfield $\tilde \Phi$ transforms as 
\ba
\delta_\zeta \tilde \Phi 
\A = \A \delta_\zeta \Phi(x'^\mu, \theta') 
\nonu
\A = \A {{\partial \Phi} \over {\partial x'^\mu}} 
\delta_\zeta x'^\mu 
+ {{\partial \Phi} \over {\partial \theta'}} 
\delta_\zeta \theta' 
\nonu
\A = \A 
{{\partial \Phi} \over {\partial x'^\mu}} 
\ \{ i \ \bar\zeta \gamma^\mu \theta' 
- \kappa^2 \ \bar\theta' \gamma^\nu \psi 
\ \bar\zeta \gamma^\rho \psi \ e_a{}^\mu \nabla_\nu e^a{}_\rho 
- i \kappa ( \delta_Q \bar\psi \gamma^\mu \theta 
+ \bar\psi \gamma^a \theta \ \delta_Q e_a{}^\mu ) \} 
\nonu
\A \A 
+ {{\partial \Phi} \over {\partial \theta'}} 
( \zeta - \kappa \ \delta_Q \psi ) 
\nonu
\A = \A \xi^\nu \left[ {{\partial \Phi} \over {\partial x'^\mu}} 
\{ \delta^\mu_\nu - i \kappa \nabla_\nu (\bar\psi \gamma^\mu \theta) \} 
+ {{\partial \Phi} \over {\partial \theta'}} 
{{\partial \theta'} \over {\partial x^\nu}} 
\right] 
\nonu
\A = \A 
\xi^\mu \partial_\mu \tilde \Phi, 
\label{tildePhi-tr}
\ea
where we have used the general covariant relation, 
$dx'^\mu = dx^\mu + \nabla_\nu (\bar\psi \gamma^\mu \theta) dx^\nu$. 
As in the case of flat spacetime \cite{IK,UZ}, 
Eq.(\ref{tildePhi-tr}) means that the components of $\tilde \Phi$ 
do not transform into each other, 
\footnote{
Note that the commutator of the homogeneous transformation (\ref{tildePhi-tr}) 
forms a closed algebra, 
$[\delta_{\zeta_1},\ \delta_{\zeta_2}] \tilde \Phi 
= \Xi^\mu \partial_\mu \tilde \Phi$.}
%
and so we expect that conditions, 
\be
{\rm components\ of}\ \tilde \Phi = {\rm constant}, 
\label{conditions}
\ee
give SUSY invariant relations which connect a L SUSY action 
with the SGM one of Eq.(\ref{SGMaction}), although rather implicitly. 

In order to materialize explicitly Eq.(\ref{conditions}) 
for a specific $N = 1$ SGM case, 
we have to investigate a (L) supermultiplet which realize 
the modified supertranslations (\ref{newstr}) in the superspace formalism. 
Therefore, as a preliminary to find the (L) supermultiplet, 
we further discuss in this letter on the realization of the modified 
superspace translations (\ref{newstr}) 
in the construction of a SUGRA-like theory 
by following the superspace formalism of WZ \cite{WZ3,WB} 
for $N = 1$ SUGRA \cite{FNFDZ}. 
In particular, we find constraints on a torsion 
and a Lorentz transformation's parameter in the superspace formalism 
to realize Eq.(\ref{newstr}) as explained later. 
We use the two-component spinor notation and superspace 
indices in Ref.\cite{WB} below, 
i.e., spinor indices are denoted by $\alpha, \beta, ... = 1, 2$ 
and $\dot \alpha, \dot \beta, ... = 1, 2$, 
while superspace indices by $M, N, ... (= \mu, \alpha, \dot \alpha)$ 
which are called as Einstein indices 
and by $A, B, ... (= a, \alpha, \dot \alpha)$ 
which are called as Lorentz indices. 
For more details of the notations and conventions see Ref.\cite{WB}. 

Let us write superspace coordinates as $z^M$ defined by 
\be
z^M = (x^\mu, \theta^\alpha, \bar\theta_{\dot \alpha}), 
\ee
and express Eq.(\ref{newstr}) as the coordinate 
transformations of $z^M$, 
\be
z'^M = z^M - \xi^M(z; e^a{}_\mu(x), \psi(x)). 
\ee
In the superspace formalism of $N = 1$ SUGRA \cite{WZ3,WB}, 
a vielbein one-form, 
\be
E^A = dz^M E_M{}^A(z) 
\ee
with a vielbein field $E_M{}^A$ (and its inverse $E_A{}^M$), 
which specifies the vierbein $e^a{}_\mu$ 
and the spin-3/2 field ($\lambda^\alpha{}_\mu$, $\bar\lambda_{\dot \alpha \mu}$), 
is introduced based on the superspace coordinates $z^M$. 
A connection one-form is also introduced as 
\be
\phi_A{}^B = dz^M \phi_{MA}{}^B 
\ee
with a connection $\phi_{MA}{}^B$. 
Note that the $\phi_{MAB}$ 
have the following property for the Lorentz indices, 
\footnote{
The $a, b$ in $(-)^{ab}$ take values 0 or 1, 
depending on whether the superspace indices are vector 
or spinor indices.}
\be
\phi_{MAB} = - (-)^{ab} \phi_{MBA}. 
\ee
Torsion and curvature two-forms, $T_A$ and $R_A{}^B$, 
are defined as the covariant derivative of 
$E^A$ and $\phi_A{}^B$, i.e., 
\ba
\A \A T_A = d E^A + E^B \phi_B{}^A = {1 \over 2} dz^M dz^N T_{NM}{}^A, 
\nonu
\A \A 
R_A{}^B = d \phi_A{}^B + \phi_A{}^C \phi_C{}^B 
= {1 \over 2} dz^M dz^N R_{NMA}{}^B. 
\ea

The transformation properties of $E_M{}^A$ and $\phi_{MA}{}^B$ 
which corresponds to the general coordinate transformation 
with the paramater $\xi_M(z)$ 
and the Lorentz transformation with a parameter $L_A{}^B(z)$ 
in the superspace formalism, are 
\footnote{
We consider the case of $L_B{}^A \not= -\xi^C \phi_{CB}{}^A$ 
in our formulation, which is a different point from Ref.\cite{WZ3,WB}. 
}
\ba
\delta E_M{}^A \A = \A - \xi^L \partial_L E_M{}^A - (\partial_M \xi^L) E_L{}^A 
+ E_M{}^B L_B{}^A 
\nonu
\A = \A - {\cal D}_M \xi^A - \xi^B T_{BM}{}^A 
+ (-)^{(m+b)c} \xi^C E_M{}^B \phi_{CB}{}^A + E_M{}^B L_B{}^A, 
\label{E-trf}
\ea
with ${\cal D}_M \xi^A = \partial_M \xi^A + (-)^{mb} \xi^B \phi_{MB}{}^A$, 
and 
\be
\delta \phi_{MA}{}^B = - \xi^L \partial_L \phi_{MA}{}^B 
- (\partial_M \xi^L) \phi_{LA}{}^B 
+ \phi_{MA}{}^C L_C{}^B - (-)^{m(a+c)} L_A{}^C \phi_{MC}{}^B 
- \partial_M L_A{}^B. 
\label{P-trf}
\ee

In the above vielbein formalism of the superspace, 
the vierbein and the spin-3/2 fields 
($e^a{}_\mu$, $\lambda^\alpha{}_\mu$, $\bar\lambda_{\dot \alpha \mu}$) 
are realized in the $\theta = \bar\theta = 0$ components of the vielbein. 
Indeed, the $E_M{}^A \mid_{\theta = \bar\theta = 0}$ is gauged into 
the form \cite{WB}, 
\be
E_M{}^A \mid_{\theta = \bar\theta = 0} 
= \pmatrix{
  e^a{}_\mu & {1 \over 2} \lambda^\alpha{}_\mu 
& {1 \over 2} \bar\lambda_{\dot \alpha \mu} \cr
  0 & \delta^\alpha{}_\beta & 0 \cr
  0 & 0 & \delta_{\dot \alpha}{}^{\dot \beta} \cr
  }, 
\ee
by using the d.o.f. of higher components of $\xi^A$ 
in ${\cal D}_\alpha \xi^A$ and $\bar{\cal D}^{\dot \alpha} \xi^A$ 
of Eq.(\ref{E-trf}). 
The inverse $E_A{}^M \mid_{\theta = \bar\theta = 0}$ is obtained as 
\be
E_A{}^M \mid_{\theta = \bar\theta = 0} 
= \pmatrix{
  e_a{}^\mu & -{1 \over 2} \lambda^\alpha{}_a 
& -{1 \over 2} \bar\lambda_{\dot \alpha a} \cr
  0 & \delta_\beta{}^\alpha{} & 0 \cr
  0 & 0 & \delta^{\dot \beta}{}_{\dot \alpha} \cr
  }. 
\ee
On the other hand, the $\theta = \bar\theta = 0$ components of the connection 
are given by \cite{WB} 
\ba
\A \A \phi_{\alpha A}{}^B(z) \mid_{\theta = \bar\theta = 0} 
= 0 = \phi^{\dot \alpha}{}_A{}^B(z) \mid_{\theta = \bar\theta = 0}, 
\nonu
\A \A 
\phi_{\mu A}{}^B(z) \mid_{\theta = \bar\theta = 0} = \omega_{\mu A}{}^B(x), 
\ea
where $\phi_{\alpha A}{}^B \mid_{\theta = \bar\theta = 0}$ 
and $\phi^{\dot \alpha}{}_A{}^B \mid_{\theta = \bar\theta = 0}$ 
have been gauged away by using the higher components 
of $L_A{}^B$ in $\partial_\alpha L_A{}^B$ 
and $\partial^{\dot \alpha} L_A{}^B$ of Eq.(\ref{P-trf}). 

Here we focus on $\delta E_M{}^A \mid_{\theta = \bar\theta = 0}$ 
which describes global SUSY transformations of 
($e^a{}_\mu$, $\lambda^\alpha{}_\mu$, $\lambda_{\dot \alpha \mu}$) 
by setting $\theta = \bar\theta = 0$ components 
of the parameter $\xi^A$ as 
\ba
\A \A \xi^a(z) \mid_{\theta = \bar\theta = 0} = 0, 
\nonu
\A \A \xi^\alpha(z) \mid_{\theta = \bar\theta = 0} = \zeta^\alpha, 
\ \ \ \bar\xi_{\dot \alpha}(z) \mid_{\theta = \bar\theta = 0} 
= \bar\zeta_{\dot \alpha}, 
\ea
As for the parameter $L_A{}^B$ which appears in Eq.(\ref{E-trf}) 
(and also in Eq.(\ref{P-trf})), 
we consider tentatively 
the case of $L_A{}^B(z) \mid_{\theta = \bar\theta = 0} \not= 0$ 
in order to realize the modified supertranslations (\ref{newstr}) 
in $\xi^a$ at ${\cal O}(\theta^1, \bar\theta^1)$. 
Indeed, this $\xi^a \mid_{(\theta^1, \bar\theta^1)}$ have to be 
chosen to preserve the gauge for $E_M{}^A \mid_{\theta = \bar\theta = 0}$, 
i.e., we consider the conditions, 
\ba
\A \A 
0 = \delta E_\alpha{}^a \mid_{\theta = \bar\theta = 0} 
= (\ - \partial_\alpha \xi^a  - \xi^\beta T_{\beta \alpha}{}^a 
- \bar\xi_{\dot \beta} T^{\dot \beta}{}_\alpha{}^a 
+ L_\alpha{}^a \ ) \mid_{\theta = \bar\theta = 0}, 
\nonu
\A \A 
0 = \delta E^{\dot \alpha a} \mid_{\theta = \bar\theta = 0} 
= (\ - \partial^{\dot \alpha} \xi^a - \xi^\beta T_\beta{}^{\dot \alpha a} 
- \bar\xi_{\dot \beta} T^{\dot \beta \dot \alpha a} 
+ L^{\dot \alpha a} \ ) \mid_{\theta = \bar\theta = 0}. 
\label{gauge}
\ea
Then we find constraints on the torsion and the Lorentz parameters 
at $\theta = \bar\theta = 0$, 
which satisfy Eq.(\ref{gauge}) and realize Eq.(\ref{newstr}) 
at the same time. 
For example, if we take values for the torsion as 
\ba
\A \A T^{\dot \beta}{}_\alpha{}^a \mid_{\theta = \bar\theta = 0} 
\ = 2i \sigma^a{}_\alpha{}^{\dot \beta} 
= T_\alpha{}^{\dot \beta a} \mid_{\theta = \bar\theta = 0}, 
\nonu
\A \A T_{\beta \alpha}{}^a \mid_{\theta = \bar\theta = 0} 
\ = 0 = T^{\dot \beta \dot \alpha a} \mid_{\theta = \bar\theta = 0} 
\label{T-constraints}
\ea
which are same as those in \cite{WZ3,WB}, 
while for the Lorentz parameters as 
\ba
\nonu
\A \A 
L_\alpha{}^a \mid_{\theta = \bar\theta = 0} 
\ = 2 \kappa^2 \sigma^\nu{}_{\alpha \dot \alpha} \bar\psi^{\dot \alpha} 
(\psi^\beta \sigma^\rho{}_{\beta \dot \beta} \bar\zeta^{\dot \beta} 
- \zeta^\beta \sigma^\rho{}_{\beta \dot \beta} \bar\psi^{\dot \beta}) 
\nabla_\nu e^a{}_\rho 
= L^a{}_\alpha \mid_{\theta = \bar\theta = 0}, 
\nonu
\A \A 
L^{\dot \alpha a} \mid_{\theta = \bar\theta = 0} 
\ = - 2 \kappa^2 
(\psi^\beta \sigma^\rho{}_{\beta \dot \beta} \bar\zeta^{\dot \beta} 
- \zeta^\beta \sigma^\rho{}_{\beta \dot \beta} \bar\psi^{\dot \beta}) 
\psi^\alpha \sigma^\nu{}_\alpha{}^{\dot \alpha} 
\nabla_\nu e^a{}_\rho 
= L^{a \dot \alpha} \mid_{\theta = \bar\theta = 0}, 
\label{L-constraints}
\ea
then Eqs.(\ref{T-constraints}) and (\ref{L-constraints}) 
lead to the $\xi^a \mid_{(\theta^1, \bar\theta^1)}$ as 
\be
\xi^a \mid_{(\theta^1, \bar\theta^1)} 
= 2 i ( \theta \sigma^a \bar\epsilon 
- \epsilon \sigma^a \bar\theta ) 
+ 2 \kappa^2 ( \theta \sigma^\nu \bar\psi 
- \psi \sigma^\nu \bar\theta )
( \psi \sigma^\rho \bar\epsilon 
- \epsilon \sigma^\rho \bar\psi ) \nabla_\nu e^a_\rho, 
\label{newstr2}
\ee
which corresponds to the modified supertranslations (\ref{newstr}). 
We note that global SUSY transformations of 
the vierbein $e^a{}_\mu$ and the spin-3/2 fields 
($\lambda^\alpha{}_\mu$, $\bar\lambda_{\dot \alpha \mu}$), 
which are derived from $\delta E_M{}^A \mid_{\theta = \bar\theta = 0}$ 
in Eq.(\ref{E-trf}), obviously depend on the NG fermion $\psi$ 
through the constraints (\ref{L-constraints}) for the Lorentz parameters. 
This fact indicates that in the linearization scenario of this letter 
the (L) supermultiplet for $N = 1$ SUSY based on the constraints 
(\ref{T-constraints}) and (\ref{L-constraints}) is a SUGRA-like multiplet 
coupled to the NG fermion $\psi$, 
in which the $\psi$ is expected to be absorbed into the londitudinal components 
of massive spin-3/2 fields, 
e.g., by means of the local spinor translations of $\psi$ in SGM. 
The coupling of NL SUSY with SUGRA has been achieved in a local SUSY invariant way 
by using a constrained chiral superfield \cite{LR} 
or by a VA superfield \cite{SW} constructed from the NG fermion. 
However, whether $N = 1$ SGM is completely equivalent to the theory 
of their works (up to the structure of auxiliary fields) has not yet known, 
for the VA model of NL SUSY \cite{VA} is equivalent to 
a scalar supermultiplet \cite{WZ1} 
and on the other hand it is also equivalent to a (axial vector) gauge supermultiplet 
\cite{WZ2} as shown in the flat spacetime case 
\cite{IK}-\cite{STT1} for $N = 1$ SUSY. 
Therefore, the relation between the theory in \cite{LR,SW} 
and $N = 1$ SGM based on the global NL SUSY transformations 
(\ref{NLSUSY-SGM}) yet to be investigated. 

Finally, we summarize our results as follows. 
In this letter, by following the systematic procedure 
\cite{IK,UZ} to linearize NL SUSY in flat spacetime, 
we have considered the application of its procedure 
to curved (SGM) spacetime for $N = 1$ SUSY 
towards the linearization of NL SUSY of Eq.(\ref{NLSUSY-SGM}) in SGM. 
Indeed, we have introduced modified superspace translations (\ref{newstr}) 
of the superspace coordinates 
and their specific coordinate transformations (\ref{newcd}) 
both depending on the NG fermion, 
and we have shown that they lead to the homogeneous transformation's law 
(\ref{tildePhi-tr}) which is important 
in the relation between L and NL SUSY. 
Furthermore, as a preliminary to prove explicitly 
that the conditions (\ref{conditions}) lead to SUSY invariant relations 
which connect a L SUSY action with the SGM one of Eq.(\ref{SGMaction}), 
we have also discussed on the realization of the modified superspace 
translations (\ref{newstr}) in the construction of a SUGRA-like multiplet 
by following the superspace formalism of WZ 
\cite{WZ3,WB} for $N = 1$ SUGRA \cite{FNFDZ}. 
In particular, we have found that constraints 
on the torsion and the Lorentz transformation parameter 
at ${\cal O}(\theta^0, \bar\theta^0)$ which satisfy the conditions (\ref{gauge}), 
e.g., Eqs.(\ref{T-constraints}) and (\ref{L-constraints}), 
realize the modified supertranslations (\ref{newstr}) 
in the parameter $\xi^a$ at ${\cal O}(\theta^1, \bar\theta^1)$ 
as shown in Eq.(\ref{newstr2}). 
A SUGRA-like action coupled to the NG fermion $\psi$ 
which is invariant under global SUSY transformations 
and the analogous mechanism to the super-Higgs one \cite{DZ} 
in the action are now under investigation 
to materialize Eq.(\ref{conditions}).

\newpage

%
\newcommand{\NP}[1]{{\it Nucl.\ Phys.\ }{\bf #1}}
\newcommand{\PL}[1]{{\it Phys.\ Lett.\ }{\bf #1}}
\newcommand{\CMP}[1]{{\it Commun.\ Math.\ Phys.\ }{\bf #1}}
\newcommand{\MPL}[1]{{\it Mod.\ Phys.\ Lett.\ }{\bf #1}}
\newcommand{\IJMP}[1]{{\it Int.\ J. Mod.\ Phys.\ }{\bf #1}}
\newcommand{\PR}[1]{{\it Phys.\ Rev.\ }{\bf #1}}
\newcommand{\PRL}[1]{{\it Phys.\ Rev.\ Lett.\ }{\bf #1}}
\newcommand{\PTP}[1]{{\it Prog.\ Theor.\ Phys.\ }{\bf #1}}
\newcommand{\PTPS}[1]{{\it Prog.\ Theor.\ Phys.\ Suppl.\ }{\bf #1}}
\newcommand{\AP}[1]{{\it Ann.\ Phys.\ }{\bf #1}}

\end{document}